\documentclass[11pt,a4paper,numbers,sort&compress]{article}
\usepackage[utf8]{inputenc}
\pdfoutput=1
\usepackage{jheppub}

\usepackage{subcaption}
\usepackage{physics}
\usepackage{amsmath}
\usepackage{color}
\usepackage{xcolor}
\usepackage{cancel}
\usepackage{amsmath}
\usepackage{amssymb}
\usepackage{slashed}
\usepackage{cancel}
\usepackage[normalem]{ulem}

\usepackage{tikz}

\usepackage{slashed}
\usepackage{cancel}
\usepackage[normalem]{ulem}

\newcommand{\nn}{\nonumber}
\newcommand{\fft}[2]{\frac{#1}{#2}}

\preprint{LCTP-24-04}

\title{The Giant Graviton Expansion from Bubbling Geometry}

\author[a]{Evan Deddo}

\author[a]{James T. Liu}

\author[a,b]{Leopoldo A. Pando Zayas}

\author[a]{Robert J. Saskowski}

\emailAdd{evdedd@umich.edu, jimliu@umich.edu, lpandoz@umich.edu, rsaskows@umich.edu}

\affiliation[a]{Leinweber Center for Theoretical Physics, 
University of Michigan, Ann Arbor, MI 48109, USA}

\affiliation[b]{The Abdus Salam International Centre for Theoretical Physics, 34014 Trieste, Italy}

\abstract{The superconformal index of half-BPS states in  ${\cal N}=4$ supersymmetric Yang-Mills with gauge group $U(N)$ admits an expansion in terms of giant gravitons,  ${\cal I}_N(q)={\cal I}_\infty(q) \sum\limits_{m=0}^\infty q^{mN}\hat{\mathcal I}_m(q)$, where $m$ is the number of giant gravitons. We derive this expansion directly in supergravity from the class of half-BPS solutions due to Lin, Lunin, and Maldacena in type IIB supergravity. The moduli space of these configurations can be quantized using covariant quantization methods. We review how this quantization leads to the graviton index, ${\cal I}_\infty(q)$, and present a modification that leads to the precise expression for the expansion in terms of giant gravitons. Our proposal provides a derivation of the giant graviton expansion directly in terms of supergravity degrees of freedom. We also comment on how to derive the expansion in terms of the effective Fermi droplet picture.}
\keywords{}

\date{\today}

\begin{document}

\maketitle

\section{Introduction}

An important aspect of precision holography is that finite $N$ corrections can be rigorously accounted for on both sides of the duality.  One place where this has shown up is in the superconformal index, which counts BPS states in a supersymmetric field theory.  Taking only a single fugacity for simplicity, the large-$N$ index generally takes on a simple form in the infinite-$N$ limit
\begin{equation}
    \mathcal I_N(q)\to\mathcal I_\infty(q),
\end{equation}
where $\mathcal I_\infty(q)$ has a clear holographic interpretation as the multi-graviton (or Kaluza-Klein graviton) index.  At infinite $N$, the index is $N$-independent.  However, finite-$N$ corrections do arise and can be understood on the field-theory side as resulting from trace relations removing states from the spectrum.

Recently, several works have presented evidence that the finite-$N$ corrections to the index can be organized in terms of a giant graviton expansion of the form \cite{Arai:2019aou,Arai:2020uwd,Imamura:2021ytr,Gaiotto:2021xce,Murthy:2022ien,Lee:2022vig,Imamura:2022aua,Liu:2022olj,Beccaria:2023zjw,Beccaria:2023hip, Lee:2023iil,Beccaria:2024vfx,Kim:2024ucf}
\begin{equation}
    \mathcal I_N(q)=\mathcal I_\infty(q)\left(1+\sum_{m=1}^\infty q^{mN}\hat {\mathcal I}_m(q)\right).
\label{eq:GGE}
\end{equation}
While some of the evidence for this expression was obtained empirically \cite{Gaiotto:2021xce} or by directly working with the matrix model representation of the field theory index \cite{Murthy:2022ien}, much of the power of this expansion comes from the holographic side, where $\hat{\mathcal I}_m(q)$ is the index for the worldvolume theory of a stack of $m$ wrapped D3-branes \cite{Arai:2019aou,Arai:2020uwd,Imamura:2021ytr}.

The giant graviton expansion, (\ref{eq:GGE}), extends to indices with multiple fugacities.  For example, the $\fft1{16}$-BPS index of $\mathcal N=4$ super-Yang-Mills with $U(N)$ gauge group admits a giant graviton expansion
\begin{equation}
    \mathcal I_N(p,q;y_a)=\mathcal I_\infty(p,q;y_a)\left(1+\sum_{n_1,n_2,n_3}(y_1^{n_1}y_2^{n_2}y_3^{n_3})^N\hat{\mathcal I}_{(n_1,n_2,n_3)}(p,q;y_a)\right),
\label{eq:1/16GGE}
\end{equation}
where $pq=y_1y_2y_3$.  Here the integers $n_a$ denote the number of wrapped D3-branes moving in the three orthogonal rotation planes related to the $S^5$.  A feature of both (\ref{eq:GGE}) and (\ref{eq:1/16GGE}) is that the prefactors $q^{mN}$ or $y_a^{n_aN}$ correspond to the classical motion of the wrapped D3-brane along $S^5$ while the giant graviton indices $\hat{\mathcal I}_m$ or $\hat{\mathcal I}_{(n_1,n_2,n_3)}$ account for the worldvolume fluctuations of the branes.  In particular, the giant graviton indices are independent of the rank $N$ of the gauge group, as $N$ only shows up in the classical contribution.

While it would be desirable to more fully investigate the properties of the $\fft1{16}$-BPS index, here we focus on the relatively simpler $\fft12$-BPS index, which is defined as
\begin{equation}
   \mathcal I_N(q)=\Tr_{\mathcal H^N_{\frac{1}{2}\mathrm{-BPS}}}(-1)^Fq^J,
\label{eq:12index}
\end{equation}
where $\mathcal H^N_{\frac{1}{2}\mathrm{-BPS}}$ is the Hilbert space of $\frac{1}{2}$-BPS states, $(-1)^F$ is the fermion number operator, and $J$ is a Cartan generator of the $SU(4)$ $R$-symmetry.  This index is easily evaluated, with the result
\begin{equation}
    \mathcal I_N(q)=\prod_{n=1}^N\frac{1}{1-q^n}=\fft1{(q)_N},
\end{equation}
where $(q)_m=\prod\limits_{j=1}^m(1-q^j)$ is the Pochhammer symbol.   As a consequence of the $q$-binomial theorem, this may be expanded as \cite{Stanley_2011}
\begin{equation}
    \mathcal I_N(q)=\mathcal I_\infty(q)\sum_{m=0}^\infty \mathcal (-1)^m\frac{q^{\frac{m(m+1)}{2}}}{(q)_m}q^{mN}.\label{eq:GGexpansion0}
\end{equation}
As first pointed out in \cite{Arai:2019aou}, the expectation is that the terms on the right-hand side correspond to the contributions of $m$ giant gravitons, which are D3-branes wrapping an $S^3\subset S^5$ and stabilized by angular momentum~\cite{McGreevy:2000cw, Hashimoto:2000zp}.  As noted in \cite{Lee:2022vig}, the $(-1)^m$ factor should be interpreted as the statement that although the $\frac{1}{2}$-BPS sector only contains traces of scalars, odd stacks of giant gravitons effectively behave as fermions.

By manipulating the Pochhammer symbol, the giant graviton expansion can be brought into the suggestive form
\begin{equation}
    \mathcal I_N(q)=\mathcal I_\infty(q)\left(1+\sum_{m=1}^\infty q^{mN}\fft1{(q^{-1})_m}\right).
\label{eq:GGexpansion}
\end{equation}
Comparison with (\ref{eq:GGE}) allows us to identify the $m$ giant graviton index as
\begin{equation}
    \hat{\mathcal I}_m(q)=\fft1{(q^{-1})_m}=\mathcal I_m(q^{-1}),
\end{equation}
which was highlighted in \cite{Gaiotto:2021xce}.  In particular, it was suggested there that $\mathcal I_m(q^{-1})$ counts the radial fluctuations of $m$ maximal giants where each quantized fluctuation carries $R$ charge $-1$.  More recently, the authors of~\cite{Eleftheriou:2023jxr,Lee:2023iil} provided further insight into the giant graviton expansion by showing that this expansion indeed arises from considering the probe limit of giant gravitons as D3-branes and semiclassical quantization around the probe solution. The goal of the present work will be to recover the giant graviton expansion of Eq.~(\ref{eq:GGexpansion}) directly from a fully back-reacted bubbling geometry.

Extracting field theory quantities and sub-structures directly from the geometry has been a long-standing goal of the AdS/CFT correspondence. Bubbling solutions are particularly powerful in this regard. Take, for example, the bubbling solutions in type IIB supergravity constructed in \cite{DHoker:2007fq} (see other important previous work~\cite{Lunin:2006xr,Yamaguchi:2006te}). These solutions are dual to Wilson loops in large representations determined by a Young tableau of order ${\cal O}(N^2)$ boxes. To evaluate the Wilson loop expectation value one uses the Gaussian matrix model whose solution is given, in the saddle point approximation, by the resolvent function; this very function then determines, through a system of nested equations, the full supergravity background~\cite{Okuda:2008px}.

The LLM bubbling AdS$_5$ background is the quintessential bubbling solution~\cite{Lin:2004nb}, whose solutions correspond to fully back-reacted giant gravitons dissolved into fluxes. Our approach will be to use the LLM solutions to directly account for the terms in the expansion \eqref{eq:GGexpansion}.  In order to do so, we will start with a classical LLM background and count quantized fluctuations on top of it.  Some recent work discussing the half-BPS index from the point of view of LLM geometries appeared in \cite{Chang:2024zqi}.

The rest of this Note is organized as follows. In Section \ref{sec:fluctuations}, we review the LLM geometries of~\cite{Lin:2004nb} and the corresponding quantization of~\cite{Grant:2005qc,Maoz:2005nk}, which we then use to recover the giant graviton expansion. In Section \ref{sec:fermi}, we discuss an alternative approach to quantization motivated by the Fermi droplet picture. Finally, in Section \ref{sec:disc}, we make some concluding remarks.

\section{Fluctuations of Giant Gravitons}\label{sec:fluctuations}

In seeking a holographic understanding of the $\fft12$-BPS giant graviton expansion, (\ref{eq:GGexpansion}), one is naturally led to the consideration of $\fft12$-BPS configurations in supergravity.  From a brane perspective, this corresponds to a stack of D3-branes wrapped on a $S^3$ in $S^5$ and orbiting in a single rotation plane corresponding to the charge $J$ highlighted by the index, (\ref{eq:12index}).  This picture of $\hat{\mathcal I}_m(q)$ as counting D3-brane fluctuations was one of the original motivations behind the giant graviton expansion \cite{Arai:2019aou,Arai:2020uwd,Imamura:2021ytr,Gaiotto:2021xce}, and was further developed in \cite{Eleftheriou:2023jxr,Lee:2023iil}.

Here we take a complementary approach and address the question of whether the index can be holographically reproduced by enumerating $\fft12$-BPS \textit{geometries}.  Such backgrounds were obtained by Lin, Lunin, and Maldacena \cite{Lin:2004nb}, and are commonly referred to (in this case) as bubbling AdS$_5$ solutions.  These LLM solutions are smooth geometries (provided boundary conditions are chosen appropriately) and can be thought of as D3-branes dissolved into fluxes much as AdS$_5\times S^5$ is obtained in the near horizon limit of a stack of D3-branes.

Since the index is essentially a counting problem, one has to address the issue of how to count LLM solutions.  At the classical level, LLM solutions are continuously deformable, so one must introduce some form of quantization in order to count states.  This was previously considered in \cite{Grant:2005qc,Maoz:2005nk} following the covariant quantization method developed in \cite{Crnkovic:1986ex,Zuckerman:1986vzu}.  While the multi-graviton index, ${\cal I}_{\infty}(q)$, was already obtained in \cite{Maoz:2005nk}, here we show how a simple modification leads to a complete expansion of the finite-$N$ index in terms of giant gravitons.

\subsection{LLM geometries}

The family of bubbling AdS$_5$ solutions are $\frac{1}{2}$-BPS solutions of type IIB supergravity which preserve an $SO(4)\times SO(4)\times\mathbb R$ symmetry~\cite{Lin:2004nb}. For such solutions, the IIB axion, dilaton, and three-form field strengths all vanish, and the metric takes the form
\begin{equation}
    \dd s^2=-h^{-2}(\dd t+V_i\dd x^i)^2+h^2(\dd y^2+\dd x^i\dd x^i)+ye^G\dd\Omega_3^2+ye^{-G}\dd\tilde\Omega_3^2,
\end{equation}
where $i=1,2$ and $\dd\Omega_3^2$ and $\dd\tilde\Omega_3^2$ are the line elements for two unit three-spheres.  Because of the isometries, the metric functions, $h$, $V_i$, and $G$ depend only on the three coordinates, $x_1$, $x_2$, and $y$.  Moreover, the complete solution including five-form flux is determined in terms of a single harmonic function, $z(x_1,x_2,y)$, which obeys the equation
\begin{equation}
    \partial_i\partial_iz+y\partial_y\qty(\frac{\partial_y z}{y})=0.
\label{eq:harmfunc}
\end{equation}
In particular, the metric functions $h$ and $G$ are determined according to
\begin{equation}
    h^{-2}=2y\cosh G,\qquad z=\frac{1}{2}\tanh G,
\end{equation}
and $V_i$ can be obtained from
\begin{equation}
    y\partial_yV_i=\epsilon_{ij}\partial_jz,\qquad y\qty(\partial_i V_j-\partial_j V_i)=\epsilon_{ij}\partial_y z.
\end{equation}
Potential singularities of the metric arise when either the first or second $S^3$ shrinks to zero size.  This occurs when $y\to0$, and LLM demonstrated that the solution is non-singular so long as $z=\pm\tfrac{1}{2}$ on the $y=0$ plane.  With these boundary conditions at $y=0$, the solution to the Laplacian, (\ref{eq:harmfunc}), is unique, and the solution is thus fully determined.

\begin{figure}[t]
    \centering
    \begin{subfigure}[t]{0.32\textwidth}
        \centering
        \leavevmode
        \raise.3cm\hbox{\includegraphics[scale=0.4]{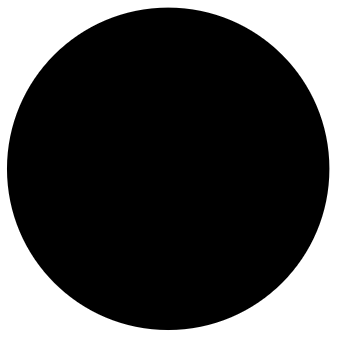}}
        \caption{}
    \end{subfigure}
    \begin{subfigure}[t]{0.32\textwidth}
        \centering
        \leavevmode
        \raise.3cm\hbox{\includegraphics[scale=0.4]{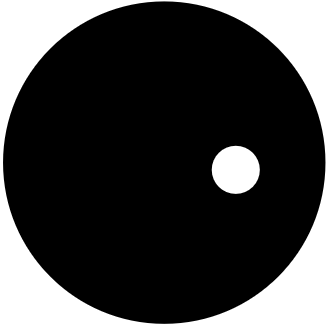}}
        \caption{}
    \end{subfigure}
    \begin{subfigure}[t]{0.32\textwidth}
        \centering
        \leavevmode
        \raise.3cm\hbox{\includegraphics[scale=0.4]{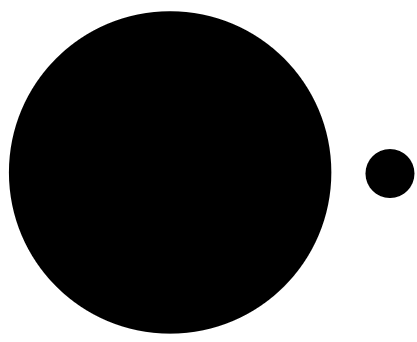}}
        \caption{}
    \end{subfigure}
    \caption{Various LLM geometries. (a) corresponds to pure $\mathrm{AdS}_5\times S^5$, (b) corresponds to asymptotically $\mathrm{AdS}_5\times S^5$ with a giant graviton present, and (c) corresponds to asymptotically $\mathrm{AdS}_5\times S^5$ with a dual giant graviton present. Note that the giants and dual giants need not be perfectly circular.}
    \label{fig:LLM}
\end{figure}

Thus, the essential point is that LLM geometries are specified by two-colorings\footnote{For graphical depiction, we will use black to color the $z=-\tfrac{1}{2}$ regions and white to color the $z=+\tfrac{1}{2}$ regions.} of the $(x_1,x_2)$-plane, which we will refer to as droplets. Let $\mathcal D$ denote the region of the $y=0$ plane where $z=-\tfrac{1}{2}$ with boundary $\partial\mathcal D$. The complement $\mathcal D^c$ has $z=+\tfrac{1}{2}$. So that the geometry is smooth, we must assume that $\partial\mathcal D$ is a smooth curve. If the droplets are of finite size, then the spacetime is asymptotically $\mathrm{AdS}_5\times S^5$. In particular, $\mathrm{AdS}_5\times S^5$ corresponds to a disk of radius $R$ in $\mathcal D$, where $R=L^2$ with $L$ being the AdS radius. Giant gravitons then correspond to a disk with some number of droplets missing, and dual giants correspond to droplets outside the disk \cite{Lin:2004nb}. See Figure \ref{fig:LLM} for examples.  Maximal giants, \emph{i.e.}, those with maximum angular momentum, correspond to a droplet in the center of the AdS disk.

While the LLM geometries are classical solutions to IIB supergravity, quantization of the self-dual five-form flux leads to a quantization of the area of $\mathcal D$ in integer units of $(2\pi)^2\ell_p^4$, where $\ell_p$ is the Planck length.  Following LLM, we introduce an effective Planck's constant, $\hbar=2\pi\ell_p^4$, in which case flux quantization takes the form
\begin{equation}
    N=\frac{1}{2\pi\hbar}\int_{\mathcal D}\dd[2]x\in\mathbb N.
\label{eq:N}
\end{equation}
Here, $N$ is identified with the flux of $F_{(5)}$ through $S^5$, or equivalently the number of D3-branes that have dissolved into fluxes.  This choice of an effective $\hbar$ is motivated by thinking of the ($x_1$,$x_2$) plane as phase space with minimum area $2\pi\hbar$.

Flux quantization also requires that each $z=+\tfrac{1}{2}$ droplet inside of $\mathcal D$ be quantized
\begin{equation}
    m_i=\frac{1}{2\pi\hbar}\int_{\mathrm{droplet}}\dd[2]x\in\mathbb N,
\label{eq:midrop}
\end{equation}
where $m_i$ is interpreted as the number of giant gravitons at position $i$. The sum $m=\sum_i m_i$ is the total number of giant gravitons. Likewise, the $z=-\tfrac{1}{2}$ droplets outside the disk satisfy a similar area quantization
\begin{equation}
    \bar m_i=\frac{1}{2\pi\hbar}\int_{\overline{\mathrm{droplet}}}\dd[2]x\in\mathbb N,
\label{eq:m}
\end{equation}
where $\bar m_i$ is interpreted as the number of dual giant gravitons at site $i$ and $\bar m=\sum_i\bar m_i$ is the total number of dual giant gravitons.

The energy $\Delta$ and angular momentum $J$ may be extracted from the asymptotic behavior of the metric. Due to the $\tfrac{1}{2}$-BPS nature of the LLM geometries, they have $\Delta=J$. These charges may be expressed as \cite{Lin:2004nb}
\begin{equation}
    \Delta=J=\frac{1}{4\pi\hbar^2}\qty[\int_{\mathcal D}\dd[2]x\,\qty(x_1^2+x_2^2)-\frac{1}{2\pi}\qty(\int_{\mathcal D}\dd[2]x)^2].
\end{equation}
Note that, while the area of $\mathcal D$ is quantized, the angular momentum, $J$, is not yet quantized, as continuous area preserving deformations of the boundary, $\partial\mathcal D$, will lead to continuous variations of $J$.  In order to obtain a quantized $J\in\mathbb N$, as one would expect for the index, we must additionally quantize the fluctuations of the moduli space of ripples on $\partial\mathcal D$.  This is what we turn to next.

\subsection{Quantization of LLM moduli space}

\begin{figure}[t]
    \centering
    \begin{subfigure}[t]{0.49\textwidth}
        \centering
        \leavevmode
        \raise.3cm\hbox{\includegraphics[scale=0.4]{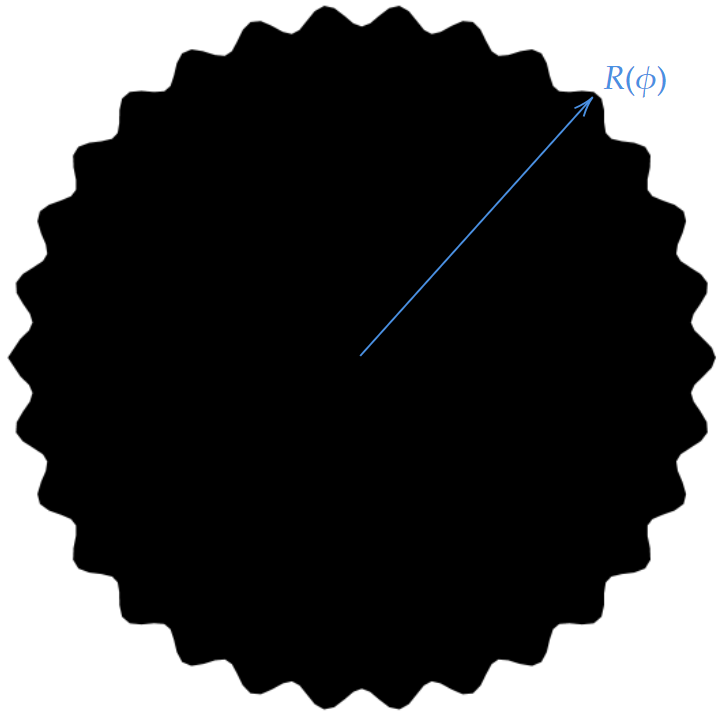}}
        \caption{}
        \label{fig:diskfluc}
    \end{subfigure}
    \begin{subfigure}[t]{0.49\textwidth}
        \centering
        \includegraphics[scale=0.4]{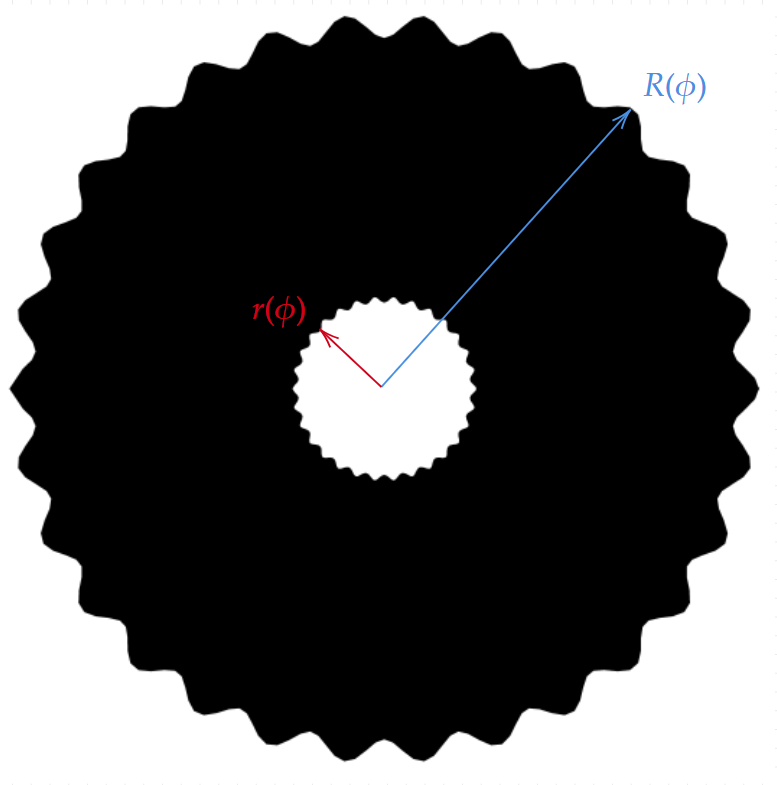}
        \caption{}
        \label{fig:giantfluc}
    \end{subfigure}
    \caption{The LLM description of $\mathrm{AdS}_5\times S^5$ is given by a disk of radius $R$. Here we schematically 
    portray fluctuations about this background.  (a) corresponds to graviton fluctuations parametrized by the curve $R(\phi)$; (b) corresponds to a maximal giant with both graviton fluctuations on the outer boundary parametrized by $R(\phi)$ and fluctuations of the maximal giant on the inner boundary parametrized by $r(\phi)$.}
    \label{fig:diskandgiant}
\end{figure}

In order to holographically reproduce the $\fft12$-BPS index, we would like to count supergravity states at fixed $N$.  This corresponds to holding the quantized area of the region $\mathcal D$ fixed according to (\ref{eq:N}), while allowing both fluctuations of the boundary, $\partial\mathcal D$, and topology change.  As observed in \cite{Lin:2004nb}, boundary fluctuations, as shown in Figure~\ref{fig:diskfluc}, correspond to graviton modes, while giant gravitons change the topology of the solution.  Maximal giants with fluctuations are depicted in Figure~\ref{fig:giantfluc}.  Classically, these fluctuations live in a continuous moduli space.  However, they were quantized in \cite{Grant:2005qc,Maoz:2005nk} using the covariant quantization method of \cite{Crnkovic:1986ex,Zuckerman:1986vzu}. 

It should be emphasized that covariant quantization only captures certain aspects of the full quantization of the geometry. In particular, it focuses on fluctuations of the moduli space, which in this case is fluctuations of the boundary, $\partial\mathcal D$.  The covariant quantization method, therefore, only captures aspects of those fluctuations and, subsequently, neglects others. However, we demonstrate that this is sufficient to describe the giant graviton expansion of the index but does not necessarily explain why.  In particular, we work in the regime where the moduli space is defined by only two functions of the contour, as in Figure~\ref{fig:giantfluc}, without including further contributions.

Before proceeding to the full giant graviton expansion, we recall the work of \cite{Maoz:2005nk} in quantizing the fluctuations of gravitons.  For the case of pure $\mathrm{AdS}_5\times S^5$, we are just considering a disk of radius $R$ in the $(x_1,x_2)$-plane.  In this case, it is convenient to introduce polar coordinates $(r,\phi)$. Then, allowing the disk to fluctuate, we parametrize the boundary of the disk by a single-valued function $R(\phi)$, as in Figure~\ref{fig:diskfluc}.  It was shown in~\cite{Maoz:2005nk} that the symplectic form for these fluctuations is given by
\begin{equation}
    \omega=\frac{1}{32\pi\hbar}\oint\dd\phi\oint\dd\tilde\phi\operatorname{Sign}(\phi-\tilde\phi)\delta\qty[R(\phi)^2]\land\delta\qty[R(\tilde\phi)^2],
\end{equation}
which leads to Poisson brackets%
\footnote{Note that the Poisson bracket $\{q_i,q_j\}=A_{ij}$ is related to the symplectic form by $\omega=(A^{-1})_{ij}\dd q_i\land \dd q_j$. This can be seen from the fact that the kernel $\mathrm{Sign}(\phi)$ is the inverse of $\delta'(\phi)$.}
\begin{equation}
    \left\{R(\phi)^2,R(\tilde\phi)^2\right\}=8\pi\hbar\delta'(\phi-\tilde\phi).\label{eq:Poisson}
\end{equation}
We must keep the area of each droplet fixed, and so we further impose the constraint
\begin{equation}
    \oint\dd\phi\,\delta\qty[R(\phi)^2]=0.
\end{equation}
In particular, we expand this as a Fourier series
\begin{equation}
    R(\phi)^2=\sum_{n\in\mathbb Z}\alpha_n e^{in\phi},\qquad \alpha_0=R^2,\qquad\alpha_{-n}=\alpha_n^*,
\end{equation}
where the $n=0$ mode is fixed by the droplet area. We may substitute this into \eqref{eq:Poisson} to get
\begin{equation}
    \{\alpha_m,\alpha_n\}=-4\hbar im\delta_{m+n},
\end{equation}
which may be quantized by promoting Poisson brackets to commutators%
\footnote{The unusual factor of $\hbar^2$ arises through the effective definition, $\hbar=2\pi\ell_p^4$, as mentioned above when connecting the droplet area to flux quantization of the IIB five-form.}
\begin{equation}
    [\alpha_m,\alpha_n]=4\hbar^2 m\delta_{m+n}.
\end{equation}

One can then  express the energy as
\begin{equation}
    \Delta = J =\frac{1}{4\pi\hbar^2}\qty[\int_0^{2\pi}\dd\phi\,\frac{R(\phi)^4}{4}-\fft\pi2 R^4]=\fft1{4\hbar^2}\sum_{n\ge1}\alpha_{-n}\alpha_n+\mathrm{const},
\end{equation}
up to a constant due to the ordering ambiguity.  Normalizing the operators according to $\alpha_m=2\hbar a_m$ then leads to $\Delta=J=\sum_{n\ge1}a_{-n}a_n$, where $[a_m,a_n]=m\delta_{m+n}$.  Thus the graviton fluctuations are quantized as a set of free bosons, and we immediately arrive at the index \cite{Maoz:2005nk}
\begin{equation}
    \mathcal I_\infty(q)=\prod_{n=1}^\infty\frac{1}{1-q^n},
\end{equation}
which is what we expect as the contribution to the index due to multi-graviton modes, as originally explained in \cite{Kinney:2005ej}.

\subsection{The giant graviton expansion}

We now turn to the giant graviton contribution to the index.  To do so, we need to sum over giant graviton topologies, which correspond to droplets removed from the interior of the LLM disk.  We thus have to consider the case of covariant quantization with disjoint boundary, $\partial\mathcal D$.  In particular, we will be interested in the case that $\partial \mathcal D$ has a set of collected components labeled by $B$ such that $\partial \mathcal D=\bigcup_{b\in B}\partial \mathcal D^{(b)}$. Assume that $\partial\mathcal D^{(b)}$ is described by a closed curved $\gamma^{(b)}(s)$ and let $\delta\gamma_\perp^{(b)}(s)$ denote the outward-directed variation of $\partial\mathcal D^{(b)}$ in the normal direction at a point $s\in\partial\mathcal D^{(b)}$. When applicable, this is related to $r(\phi)$ by
\begin{equation}
    \frac{\dd s}{r(\phi)\dd\phi}=\frac{\delta r}{\delta\gamma_\perp^{(b)}}.
\end{equation}
This then has a symplectic form \cite{Maoz:2005nk}
\begin{equation}
    \omega=\frac{1}{8\pi\hbar}\sum_{b\in B}\oint_{\gamma^{(b)}}\dd s\oint_{\gamma^{(b)}}\dd\tilde s\operatorname{Sign}(s-\tilde s)\,\delta\gamma_\perp^{(b)}(s)\land\delta\gamma_\perp^{(b)}(\tilde s),
\end{equation}
and correspondingly satisfies a Poisson bracket 
\begin{equation}
    \left\{\delta\gamma_\perp^{(b)}(s),\delta\gamma_\perp^{(\tilde b)}(\tilde s)\right\}=2\pi\hbar\delta'(s-\tilde s)\delta_{b\tilde b}.\label{eq:poisson2}
\end{equation}
This will be useful to us in the case of giant gravitons, which have multiple droplet boundaries. Importantly, different droplet boundaries are completely decoupled.
This is subject to the constraint of droplet area quantization
\begin{equation}
    \oint_{\gamma^{(b)}}\dd s\,\delta\gamma_\perp^{(b)}(s)=0,\label{eq:constraint}
\end{equation}
which specifies symplectic sheets in the moduli space of droplets.

In principle, one may wish to consider multiple giant gravitons, corresponding to multiple droplets removed from the LLM disk.  However, here we restrict our focus to maximal giants only.  These maximal giants are all overlapping and centered at the origin of the LLM plane, and hence yield a configuration of the form shown in Figure~\ref{fig:giantfluc}, with the area of the central hole related to the number of maximal giants according to (\ref{eq:midrop}).  To be specific, consider a configuration of $m$ maximal giant gravitons. Then \eqref{eq:N} and \eqref{eq:midrop} tell us that we must have
\begin{equation}
    N=\frac{R^2-r^2}{2\hbar},\qquad m=\frac{r^2}{2\hbar}\label{eq:maxcharge}.
\end{equation}
We may parametrize the outer boundary by a curve $R(\phi)$ and the inner boundary by a curve $r(\phi)$, as shown in Figure \ref{fig:giantfluc}, such that
\begin{align}
    R(\phi)^2&=\sum_{n\in\mathbb Z}\alpha_n e^{in\phi},\qquad \alpha_0=R^2,\qquad\alpha_{-n}=\alpha_n^*,\nonumber\\
    r(\phi)^2&=\sum_{j\in\mathbb Z}\beta_j e^{ij\phi},\qquad \beta_0=r^2,\qquad \beta_{-j}=\beta_j^*.
\end{align}
This parametrization automatically satisfies the area-preserving constraint, \eqref{eq:constraint}. Moreover, after promoting Poisson brackets to commutators, \eqref{eq:poisson2} requires that the modes satisfy
\begin{equation}
    [\alpha_m,\alpha_n]=4\hbar^2 m\delta_{m+n},\qquad [\beta_m,\beta_n]=4\hbar^2 m\delta_{m+n},\qquad [\alpha_m,\beta_n]=0.
\end{equation}
Using \eqref{eq:maxcharge}, the maximal giants then have charges
\begin{equation}
    \Delta=J=\frac{1}{4\pi\hbar^2}\qty[\int_0^{2\pi}\dd\phi\,\frac{R(\phi)^4-r(\phi)^4}{4}-\fft\pi2\qty(R^2-r^2)^2]=mN+\sum_{n\ge 0}a_{-n}a_n-\sum_{j\ge 0}b_{-j}b_j,
\end{equation}
where $a_m=\alpha_m/(2\hbar)$ and $b_j=\beta_j/(2\hbar)$ are the normalized operators satisfying the free boson commutation relations
\begin{equation}
    [a_m,a_n]=m\delta_{m+n},\qquad[b_m,b_n]=m\delta_{m+n}.
\end{equation}
In order to interpret this result, the first term on the right corresponds to $J=mN$, which is the `classical' angular momentum of $m$ maximal giants, and the $a_n$ fluctuations correspond to quantized multi-graviton fluctuations on the outer boundary of the LLM disk, giving $\mathcal I_\infty(q)=1/(q)_\infty$ as before.

Now consider the contribution of the giant graviton fluctuations, $\hat {\mathcal I}_m(q)$.  A simple application of counting the $b_j$ modes would lead to the formal expression $1/(q^{-1})_\infty$.  However, unlike for the case of the outer boundary, we may not make the approximation that the inner boundary is large enough to support arbitrary $b_j$ fluctuations, as the size of the inner boundary is associated with $\hbar$. It may be observed that $b_{j}$ changes $J$ by $j$, and so increases $r^2$ by $2\hbar j$. Consequently, denoting the $j^{\mathrm{th}}$ occupation number by $n_j$, the inner radius is
    \begin{equation}
        r(\phi)^2=2\hbar\qty(m+\sum_jn_j e^{ij\phi}).
    \end{equation}
If $\sum_j n_j>m$, then the radius squared can go negative. To avoid this, we must cut off the sum of occupation numbers at $m$. The coefficient of each term $q^{-n}$ in the expansion of $\hat {\mathcal I}_m(q)$ should therefore count the sets of occupation numbers satisfying
    \begin{equation}
        \sum_{j=0}^{\infty} n_j j = n,
    \end{equation}
with the additional constraint on the total sum, $\sum_jn_j\le m$. This counting is precisely the number of partitions of $n$ into at most $m$ parts. By a standard result from the theory of partitions, the same result is obtained by counting the number of partitions of $n$ with no part greater than $m$. Therefore,
    \begin{equation}
        \hat {\mathcal I}_m(q) =\prod_{j=1}^m\frac{1}{1-q^{-j}}=\fft1{(q^{-1})_m}=\mathcal I_m(q^{-1}),
    \end{equation}
giving the full index
\begin{equation}
    \mathcal I_N(q)=\mathcal I_\infty(q)\sum_{m=0}^\infty \mathcal I_m(q^{-1})q^{mN}.\label{eq:GGexpansion2}
\end{equation}
In particular, we see that \eqref{eq:GGexpansion2}, derived from quantizing the LLM description, matches the giant graviton expansion \eqref{eq:GGexpansion}.

\section{The Fermi Droplet Picture}\label{sec:fermi}

While we have focused on semi-classical quantization of bubbling geometries, one may take a complementary approach to $\fft12$-BPS states.  In particular, LLM solutions are dual to a subsector of chiral primaries in $\mathcal N=4$ super Yang-Mills with conformal dimension $\Delta$ and $U(1)$ $R$-charge $J$ satisfying $\Delta=J$ \cite{Lin:2004nb}. These admit a description in terms of free fermions in a harmonic oscillator potential~\cite{Corley:2001zk,Berenstein:2004kk}. In particular, the ``droplets'' in the $(x_1,x_2)$-plane in the supergravity description precisely correspond with droplets in the free fermion phase space. We now consider the effective picture of quantizing this fermion liquid.

\begin{figure}[t]
    \centering
    \includegraphics[scale=0.5]{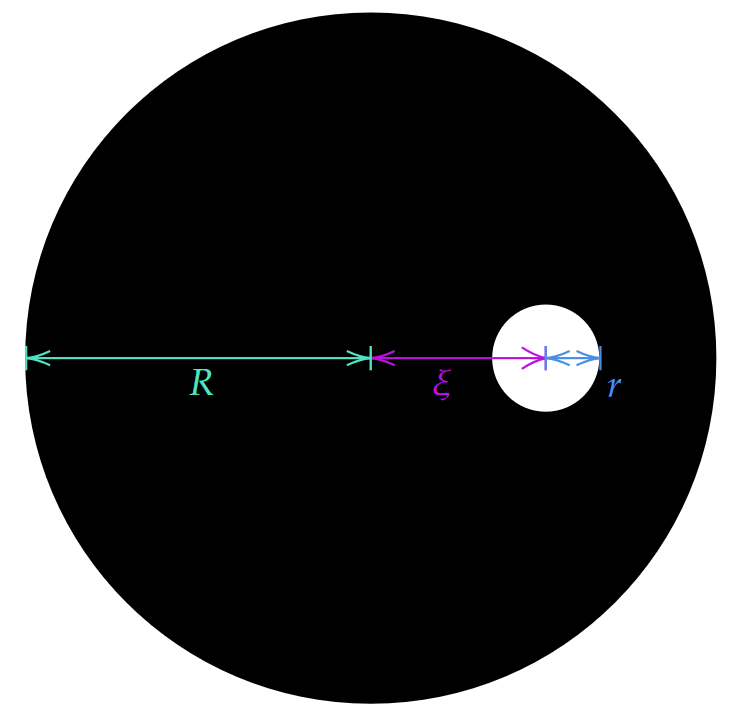}
    \caption{Schematic depiction of giants in the LLM droplet picture. When $\xi=0$, these become maximal giants.}
    \label{fig:nonmax}
\end{figure}

Consider a droplet of $m$ (not necessarily maximal) giant gravitons. Then this will have charges
\begin{equation}
    \Delta=J=\frac{r^2}{4\hbar^2}(R^2-r^2-\xi^2),
\end{equation}
where $\xi$ denotes the distance from the origin of the AdS disk, $R$ denotes the radius of the AdS disk, and $r$ denotes the radius of the giant(s) (see Figure \ref{fig:nonmax}).  Due to flux quantization, this may be recast as
\begin{equation}
    \Delta=J=m(N-p)=mp',
\end{equation}
where $p=\xi^2/2\hbar$ is the quantization of $\xi^2$, and $p'=1,...,N$ is a convenient choice of angular momentum quantization\footnote{Note that we omit $p'=0$. This corresponds to pure $\mathrm{AdS}_5\times S^5$, so we exclude it to avoid overcounting.}. Then, we have a counting problem of picking occupation numbers $n_p$ for the $N$ angular momentum levels. We can already see that this gives us the index
\begin{equation}
    \mathcal I_N(q)=\prod_{p=1}^N\sum_{n_p=0}^\infty q^{pn_p}=\prod_{p=1}^N\frac{1}{1-q^p}.
\end{equation}
Of course, we would like to get an expansion in the number of giant gravitons, and so we want to find the contribution from $m$ giants. This corresponds to imposing the constraint
\begin{equation}
    n_1+\cdots+n_N=m,\label{eq:occuConstraint}
\end{equation}
and then summing over $m$. It is not hard to see that this is just the $q$-analog of the classic balls and bins problem. Thus, we must put $m$ giant gravitons into $N$ angular momentum levels, which can be seen to have the solution
\begin{equation}
    \mathcal I_N(q)=\sum_{m=0}^\infty \begin{bmatrix}
        N+m-1\\m
    \end{bmatrix}_q q^m,\label{eq:GGexpansion3}
\end{equation}
where the brackets denote the $q$-binomial coefficients
\begin{equation}
    \begin{bmatrix}
        r\\s
    \end{bmatrix}_q\equiv\frac{(1-q^r)(1-q^{r-1})...(1-q^{r-s+1})}{(1-q)(1-q^2)...(1-q^s)}.
\end{equation}
As a sanity check, we may observe that the $q$-analog of the negative binomial theorem states that \cite{Stanley_2011}: 
\begin{equation}
    \sum_{m=0}^\infty \begin{bmatrix}
        N+m-1\\m
    \end{bmatrix}_q q^m=\prod_{p=0}^N\frac{1}{1-q^p}.\label{eq:binsballs}
\end{equation}
However, we would like to connect this with the expansion \eqref{eq:GGexpansion}. Our first observation is that term by term, the two series are distinct. We get
\begin{align}
    \mathcal I_N(q)=&1\nonumber\\
    &+\frac{q}{1-q}-\frac{q^{1+N}}{1-q}\nonumber\\
    &+\frac{q^2}{(1-q)(1-q^2)}-\frac{q^{2+N}}{(1-q)^2}+\frac{q^{3+N}}{(1-q)(1-q^2)}\nonumber\\
    &+\frac{q^3}{(1-q)(1-q^2)(1-q^3)}-\frac{q^{3+N}}{(1-q)^2(1-q^2)}+\frac{q^{4+2N}}{(1-q)^2(1-q^2)}-\frac{q^{6+3N}}{(1-q)(1-q^2)(1-q^3)}\nonumber\\
    =&\qty(1+\frac{q}{1-q}+\frac{q^2}{(1-q)(1-q^2)}+\frac{q^3}{(1-q)(1-q^2)(1-q^3)}+...)\nonumber\\
    &\times\qty(1-\frac{q^{1+N}}{1-q}+\frac{q^{3+N}}{(1-q)(1-q^2)}-\frac{q^{6+3N}}{(1-q)(1-q^2)(1-q^3)}+...)
\end{align}
The first term in parentheses we expect to be $\mathcal I_\infty(q)$ and the second term appears to be the coefficients in the expansion \eqref{eq:GGexpansion}. More formally, we may observe that \eqref{eq:GGexpansion3} can be rewritten as
\begin{equation}
    \mathcal I_N(q)=\sum_{m=0}^\infty \frac{(q^N;q)_m}{(q)_m}q^m=\sum_{m=0}^\infty\sum_{j=0}^m\frac{q^{m-j}}{(q)_{m-j}}\frac{(-1)^jq^{\frac{j(j+1)}{2}}}{(q)_j}q^{jN},
\end{equation}
where the second equality follows from the $q$-binomial theorem and we have made use of the Pochhammer symbols
\begin{equation}
    (a,q)_m\equiv\prod_{k=0}^{m-1}\frac{1}{1-aq^k},\qquad (q)_m\equiv(q,q)_m.
\end{equation}
Applying the discrete version of Fubini's theorem, we may swap the order of the sums to get
\begin{equation}
    \mathcal I_N(q)=\sum_{j=0}^\infty\sum_{m=j}^\infty\frac{q^{m-j}}{(q)_{m-j}}\frac{(-1)^jq^{\frac{j(j+1)}{2}}}{(q)_j}q^{jN}.
\end{equation}
It is then easy to do the sum over $m$ using the $q$-expansion
\begin{equation}
    \mathcal I_\infty(q)=\frac{1}{(q)_\infty}=\sum_{k=0}^\infty\frac{q^{k}}{(q)_{k}},
\end{equation}
to get that
\begin{equation}
    \mathcal I_N(q)=\mathcal I_\infty(q)\sum_{j=0}^\infty \frac{(-1)^jq^{\frac{j(j+1)}{2}}}{(q)_j}q^{jN},
\end{equation}
which is precisely the expansion \eqref{eq:GGexpansion0}. However, in contrast to the preceding section, $j$ is not directly interpreted as the number of giant gravitons (although it is related).

\subsection{Connection to deformation quantization}

Let us now connect our result with that of \cite{Chang:2024zqi}. The authors used deformation quantization to obtain a Hamiltonian
\begin{equation}
    H=\sum_{m=0}^{\infty}c_m\qty(m+\frac{1}{2})-\frac{N^2}{2},\qquad \sum_{m=0}^{\infty} c_m=N,\qquad c_m\in\{0,1\},
\end{equation}
which is just the Hamiltonian of $N$ free fermions in a harmonic oscillator potential. This is easily seen to be equivalent to
\begin{equation}
    H=\sum_{i=1}^N\qty(f_i+\frac{1}{2})-\frac{N^2}{2},\qquad 0\le f_1<f_2<...<f_N<\infty,
\end{equation}
which may then be mapped onto our Fermi droplet picture by identifying \cite{Suryanarayana:2004ig}
\begin{equation}
    n_N=f_1,\qquad n_{N-i}=f_{i+1}-f_i-1,\qquad i=1,2,...,N-1,
\end{equation}
which precisely reproduces our earlier Hamiltonian
\begin{equation}
    H=\sum_{p=1}^N pn_p.
\end{equation}
The counting \eqref{eq:occuConstraint} then corresponds to fixing
\begin{equation}
    f_N=N+m-1.
\end{equation}

\begin{figure}
    \centering
    \begin{subfigure}[t]{0.32\textwidth}
        \centering
        \leavevmode
        \raise.3cm\hbox{\includegraphics[scale=0.4]{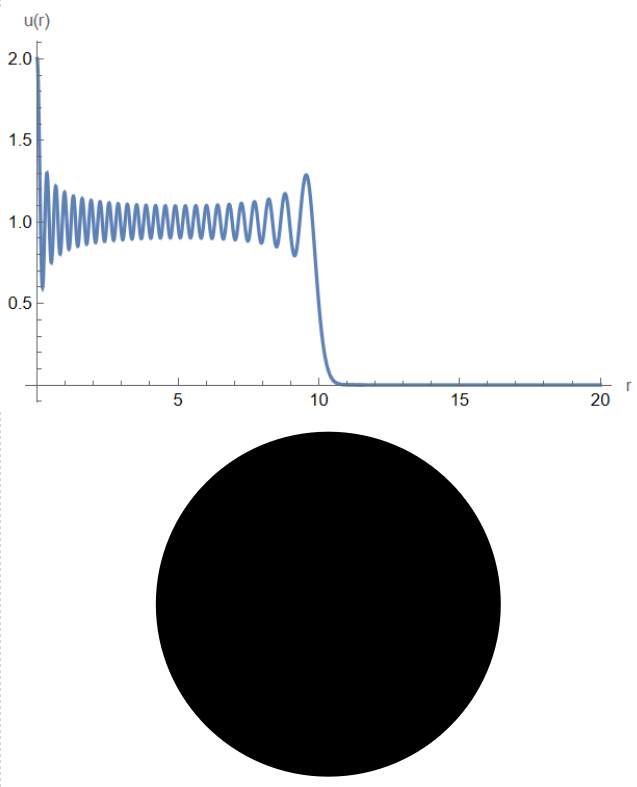}}
        \caption{}
    \end{subfigure}
   \begin{subfigure}[t]{0.32\textwidth}
        \centering
        \leavevmode
        \raise.3cm\hbox{\includegraphics[scale=0.4]{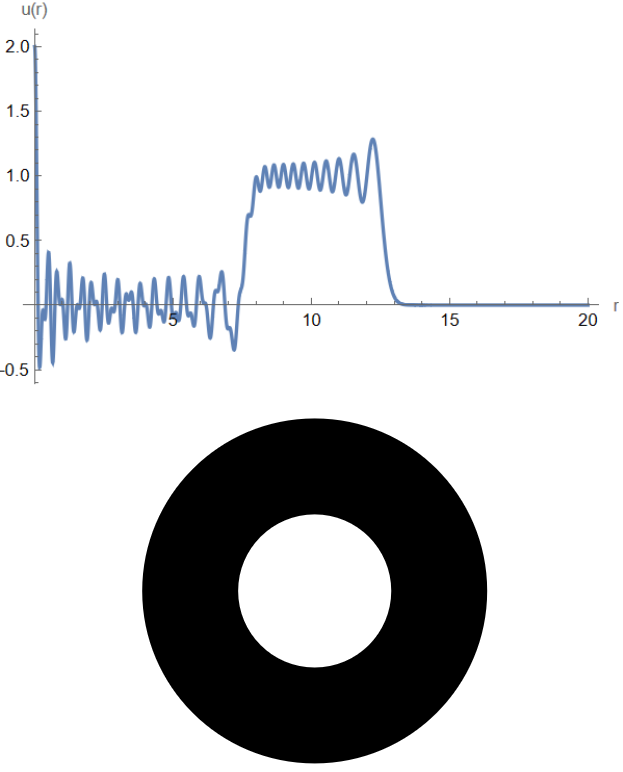}}
        \caption{}
    \end{subfigure}
    \begin{subfigure}[t]{0.32\textwidth}
        \centering
        \leavevmode
        \raise.3cm\hbox{\includegraphics[scale=0.4]{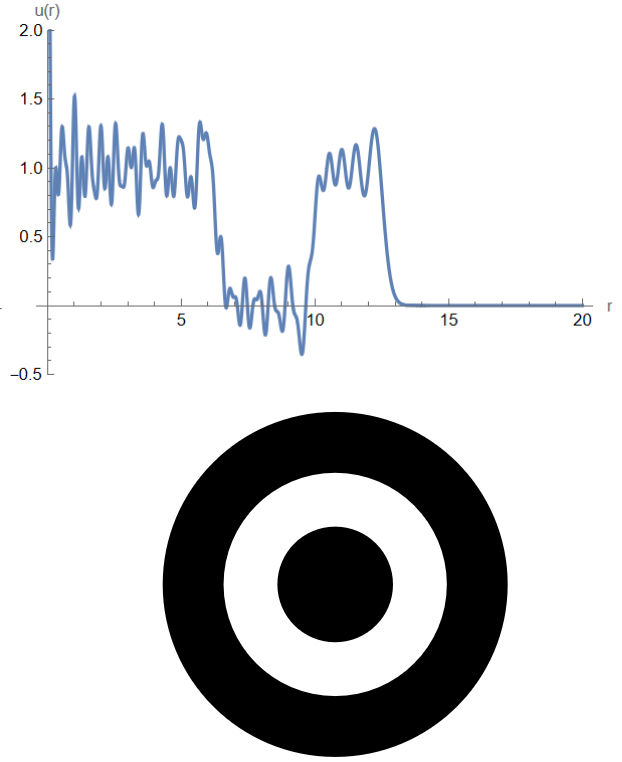}}
        \caption{}
    \end{subfigure}
    \caption{Plots of $u=\frac{1}{2}-z$ as a function of $r$ and their corresponding approximate classical interpretation in the LLM moduli space. We have set $\hbar=1$ and $N=50$ for visualization. (a) corresponds to $c_1=c_2=\cdots=c_{50}=1$ with all other $c_n$ vanishing. This is naturally identified with $\mathrm{AdS}_5\times S^5$. (b) corresponds to $c_1=\cdots=c_{30}=0$, $c_{31}=\cdots=c_{80}=1,$ and all other $c_n$ vanishing. This is naturally identified with a maximal giant. (c) corresponds to the case $c_1=\cdots=c_{20}=1$, $c_{21}=\cdots=c_{50}=0$, $c_{51}=\cdots=c_{80}=1$, and all other $c_n$ vanishing. This is interpreted as a ``new geometry.''}
    \label{fig:defQuant}
\end{figure}

However, it is important to note that the geometric interpretation is slightly different. The coloring of the $(x_1,x_2)$-plane in \cite{Chang:2024zqi} is given by
\begin{equation}
    z(x_1,x_2)=\frac{1}{2}-\sum_{n=0}^\infty c_n\phi_n(x_1,x_2),\qquad \phi_n(x_1,x_2)=2(-1)^ne^{-r^2/\hbar}L_n\qty(\frac{2r^2}{\hbar}).\label{eq:coloring}
\end{equation}
In particular, $\phi_n$ is independent of the angle $\phi$, so it cannot have the interpretation of fluctuations of giant gravitons nor as non-maximal giants displaced from the center. Rather, states of the form
\begin{equation}
    c_{n<n_0}=0,\qquad c_{n_0\le n\le n_1}=1,\qquad c_{n>n_1}=0,
\end{equation}
correspond to (non-fluctuating) maximal giant gravitons. The rest correspond to ``new geometries'' in the LLM language.  In \cite{Balasubramanian:2005mg}, the interpretation given was that classical LLM geometries should be interpreted just as an effective IR description of the
microphysics and that the classical metric loses meaning once the Planck
scale is reached. Here, having quantized the geometry, we can directly see the implications of small oscillations on the gravitational physics. The classical moduli space of these solutions roughly looks like a disk with some number of rings cut out. For example, see Figure \ref{fig:defQuant}.
Hence, this should be interpreted as an expansion in ``new quantum geometries.'' It is important to emphasize the inherent fuzziness of the colorings in \eqref{eq:coloring}. This reflects the fact that the classical geometry breaks down when quantizing, and we are left with a quantum geometry. However, classically, there is no distinction (at least in terms of the charges) between a non-maximal giant and a non-maximal giant smeared into a ring, so it is natural that the two pictures give the same expansion.

\subsection{Dual giants}
Giants and dual giants are expected to be related by the analog of particle-hole duality~\cite{Suryanarayana:2004ig}. Consider a droplet of radius $r$ of $\bar m$ dual giants at a distance $\xi$ away from the center of the AdS disk of radius $R$. We may quantize this configuration in precisely the same way as we did for giants, and we get charges
\begin{align}
    N&=\frac{R^2+r^2}{2\hbar},\qquad \bar m=\frac{r^2}{2\hbar},\qquad p=\frac{\xi^2}{2\hbar},\nn\\
    \Delta&=J=\frac{r^2}{4\hbar^2}\qty(\xi^2-R^2)=\bar m\qty(p+m-N)=\bar m p',
\end{align}
where $p'=1,2,3,...$. Note that dual giants have no bound on their angular momentum, but we may have at most $N$ dual giants due to the stringy exclusion principle. Hence, we get
\begin{equation}
    \mathcal I_N(q)=\prod_{\bar m=1}^N\frac{1}{1-q^{\bar m}},
\end{equation}
which is a product over the number of dual giants. Likewise, we may expand this as a series expansion in $J=j$ as
\begin{equation}
    \mathcal I_N(q)=\sum_{j=0}^\infty \begin{bmatrix}
        N+j-1\\j
    \end{bmatrix}_q q^j,
\end{equation}
which is precisely the same as for giant gravitons but with $m$ and $J$ switched.

\section{Discussion}\label{sec:disc}
We have accounted for the contributions in the giant graviton expansion \eqref{eq:GGexpansion} by considering fully back-reacted bubbling geometries. We have also provided an alternative counting from the perspective of the free Fermi liquid. It is somewhat puzzling in our picture that the counting restricts to maximal giants; a similar restriction was also implemented in two recent derivations of the giant graviton expansion in the probe approximation \cite{Lee:2023iil,Eleftheriou:2023jxr}. In the quantization framework that we work in, one possible resolution is to observe that the $\beta_{-1}$ mode acts as a sort of translation away from the origin, so ignoring non-maximal giants would be consistent with not double counting configurations.

It would be interesting to see if this story also extends to the case of bubbling configurations in M-theory. The solutions are still determined by two-colorings of a plane representing choices of boundary conditions, now corresponding to the solution of a particular Toda equation \cite{Lin:2004nb}. Depending on how these are chosen, the solutions correspond to either bubbling $\mathrm{AdS}_4\times S^7$ or bubbling $\mathrm{AdS}_7\times S^4$.  The $\mathrm{AdS}_4$ case corresponds to M5 branes wrapping $S^5\subset S^7$ and is dual to $U(N)_k\times U(N)_{-k}$ ABJM theory at level $k=1$ \cite{Beccaria:2023cuo}. The $\mathrm{AdS}_7$ case corresponds to M2 branes wrapping $S^2\subset S^4$  and is dual to the $\mathcal N=(2,0)$ theory \cite{Beccaria:2023sph}. The corresponding giant graviton expansions are known  to the first few orders
\begin{align}
    \mathcal I^{\mathrm{ABJM}}_N(q)&=\mathcal I^{\mathrm{ABJM}}_\infty(q)\qty[1-\frac{N^3}{6}\qty(\frac{q}{(q)_\infty^7}+\mathcal O\qty(N^{-1}))q^N+\mathcal O\qty(q^{2N})],\nn\\
    \mathcal I^{\mathcal N=(2,0)}_N(q)&=\mathcal I^{\mathcal N=(2,0)}_\infty(q)\qty[1-\frac{q}{(1-q)^2}q^N+\frac{2q^3}{(1-q^2)^2(1-q)^2}q^{2N}+\mathcal O\qty(q^{3N})],
\end{align}
The $N^3$ factor seems to break the typical pattern \eqref{eq:GGE} but it is still a giant graviton expansion in the sense of expanding in contributions of wrapped branes. It would be interesting to see if the same phenomena occur in those cases as well. Here, the graviton indices are given by
\begin{align}
    \mathcal I^{\mathrm{ABJM}}_\infty(q)&=\mathrm{PE}\qty[\frac{-q^8+q^4+\left(q^2+q+1\right)^4
   \left(q^4-1\right)^2-1}{\left(q^4-1\right)^4}],\nn\\
    \mathcal I^{\mathcal N=(2,0)}_\infty(q)&=\prod_{n=1}^\infty \frac{1}{(1-q^n)^n},
\end{align}
where PE denotes the plethystic exponential. However, the LLM energy and angular momentum have not been computed, nor has the symplectic structure of phase space, so we leave this to future work. 

\section*{Acknowledgements}
We are particularly grateful to Ji Hoon Lee for various insightful comments. This work is partially supported by the U.S. Department of Energy under grant DE-SC0007859. This research was supported in part by grant NSF PHY-2309135 to the Kavli Institute for Theoretical Physics (KITP).

\bibliographystyle{JHEP}
\bibliography{B-BH.bib}
\end{document}